\RequirePackage{fix-cm}
\documentclass{svjour3}                     
\smartqed  
\usepackage{graphicx}
\usepackage{amsmath}
\usepackage{amssymb}

\usepackage{mathptmx}      

\usepackage{color}

\newcommand {\be}  {
\begin{equation}
}
\newcommand {\ee}  {
\end{equation}
}
\newcommand {\bea} {
\begin{eqnarray}
}
\newcommand {\eea} {
\end{eqnarray}
}

\newcommand{\uj}{u^{(j)}}
\newcommand{\ujs}{u^{(j)*}}
\newcommand{\rv}{\vec{r}}
\newcommand{\Rv}{\vec{R}}
\newcommand{\Aj}{A^{(j)}}
\newcommand{\Ajs}{A^{(j)*}}
\newcommand{\epsnew}{\tilde{\epsilon}}
\newcommand{\khj}{\hat{k}^{(j)}}
\newcommand{\kb}{k_\mathrm{B}}
\newcommand{\kv}{\vec{k}}
\newcommand{\kh}{\mathbf{\hat{k}}}

\begin{document}

\title{Modeling of grain boundary dynamics using amplitude equations
}


\author{Claas H\"uter         \and
             J\"org Neugebauer \and
             Guillaume Boussinot \and
             Bob Svendsen \and
             Ulrich Prahl \and
             Robert Spatschek
}


\institute{C. H\"uter \and J. Neugebauer \and R. Spatschek \at
              Computational Materials Design, Max-Planck Institute for Iron Research, 40237 D\"usseldorf, Germany \\
              Tel.: +49-211-6792-820\\
              Fax: +49-211-6792-465\\
              \email{hueter@mpie.de}
           \and
           G. Boussinot \at
              Access e.V., RWTH 52072 Aachen, Germany \\
              Peter-Gr\"unberg Institute, Research Center 52425 J\"ulich, Germany
            \and
            B. Svendsen \at
            Microstructure Physics and Alloy Design, Max-Planck Institute for Iron Research, 40237 D\"usseldorf, Germany \\
            Material Mechanics, 52062 RWTH Aachen, Germany
            \and
            U. Prahl \at
            Department of Ferrous Metallurgy,
            RWTH 52056 Aachen, Germany
}

\date{Received: date / Accepted: date}

\maketitle

\begin{abstract}
We discuss the modelling of grain boundary dynamics within an amplitude equations description, which is derived from classical density functional  theory or the phase field crystal model.
The relation between the conditions for periodicity of the system and coincidence site lattices at grain boundaries is investigated.
Within the amplitude equations framework we recover predictions of the geometrical model by Cahn and Taylor for coupled grain boundary motion, and find both $\langle100\rangle$ and $\langle110\rangle$ coupling.
No spontaneous transition between these modes occurs due to restrictions related to the rotational invariance of the amplitude equations.
Grain rotation due to coupled motion is also in agreement with theoretical predictions.
Whereas linear elasticity is correctly captured by the amplitude equations model, open questions remain for the case of nonlinear deformations.
\end{abstract}

\section{Introduction}
\label{intro}
The phase field method has a long track of remarkable success in various branches of applied and theoretical physics and engineering. 
Generally speaking, it is an approach tailored to interfacial pattern formation problems, which arise in various classical phase transformations that are formulated as free boundary problem. 
The phase field method introduces an additional variable to describe the phase state. 
This variable, the phase field or order parameter, yields a smooth transition between the phases on an artificial length scale \cite{LangerLesHouchesLectures}. 
From a historical perspective, next to Landau theory \cite{LandauZhEkspTeorFiz7_19_1937} the work of Cahn \cite{ActaMetallCahn1960} on discrete and diffusive interfaces in phase transitions, which  introduces the scaling for the length scale associated with a finite interface thickness, is probably the most influential preliminary work. 
These publications are fundamental to the seminal developments by Fix \cite{PF1st_Fix1983} and Langer \cite{Langer:1986fk}, who presented the method originally. 
Right from the start, the new approach led to several milestones of solid state simulation. 
We just name Hillert's discrete model for spinodal decomposition \cite{HillertThesis1956}, the continuous model of  Cahn and Hilliard for the same problem \cite{HilliardCahn1958,HiiliardCahn1959}, which uses the alloy concentration as order parameter, and Khachaturyan's theory of micro-elasticity \cite{Khachaturyan1983Book} which is fundamental to a group of phase-field models which focus on the application to microstructure evolution \cite{WangKhachaturyanMaterSciEngA2006,ChenReview,Wang:2010aa}. 

Among this wide range of interesting topics, very prominent examples of successful combinations with complementary methods are the solidification of pure materials or alloys and various solid-state transformations, see the reviews of Karma and Boettinger et al. \cite{KarmaReview,BoettingerReview}. 
Especially the combination with boundary integral descriptions \cite{BoettingerReview,Brener:2009qo,Hueter:2012syn,PRE_83_020601_Boussinotetal_2011,PRE83_050601_Hueteretal_2011} proved to be very successful. 
Together, a comprising understanding of the fundamental aspects of such phase transitions, covering both the aspects of stability and dynamics as well as asymptotic behaviour and basic scaling laws, could be obtained. 

As the phase field community grew with increasing success of the method, the theory evolved mainly in two branches, the order-parameter and the indicator-field interpretation, see also the reviews of Chen and Steinbach \cite{ChenReview,steinbach}.
The indicator-field models assign to thermodynamically distinguishable phases the material data and are often used for coupled dynamics of e.g. elasticity and diffusion, 
while the physical order-parameter models are mostly used to describe order-disorder transitions, phase separations or martensitic transformations \cite{ActaMat46_2983_1998_WangYetal,ActaMat47_1995_1999_RubinGetal,IntJSolidsStruct37_2000_DreyerWetal,PhaseTransitions69_1999_KerrWCetal}. While the developed phase field models could be modified to describe even atomistic scale effects, like premelting \cite{PhysRevE.81.051601,Pavan2014PRE}, the phase field crystal (PFC) method which was introduced quite recently by Elder et al.~\cite{PhysRevLett.88.245701} provides a natural description of such effects. 
Specifically, the phase field crystal theory describes the phenomena on atomic length and diffusive time scales. The former naturally yields elastic and plastic deformation, and the latter allows simulations on time scales much larger than comparable atomic methods. The PFC model was shown to be consistent with predictions for the grain boundary energy and misfit dislocations in epitaxial growth, showing the capacity to describe atomistic scale phenomena. 
The remaining drawback of the PFC method is the required spatial discretisation on atomistic or even sub-atomic length scales. 

In this article we present results on an approach for materials science modelling based on {\em amplitude equations} which compensates this limitation of the PFC method. 
Amplitude equations are well known in pattern formation modelling, especially in hydrodynamics \cite{RevModPhys.65.851}. 
The transfer to cubic crystal systems \cite{Spatschek:2010fk,Wu:2006uq} showed the potential of this elegant 
and computationally efficient method. 
The amplitude equations model might be considered as ``phase field with atoms'', while the involved coarse-graining process allows a quantitative link to atomistic modelling methods like molecular dynamics and classical density functional theory \cite{ISI:A1981LE27100066,ISI:A1987K583500046,Singh1991351,ISI:A1984SD76100042,ISI:A1996TY72800041,ISI:A1996VM55400040}.
In combination with recent studies on premelting and atomistic effects in grain boundary melting \cite{kar13,MSMSE_22_034001_2014_HueterNguyenSpatschekNeugebauer,CHueterPRB2014}, this demonstrates the capacity of amplitude equations to provide insights which were previously not accessible by continuum approaches.
At the same time it offers the possibility to describe large scale coarsening phenomena with elastic effects, which were previously studied using scaling analyses \cite{Brener:2000fk}.
In particular we investigate in the present paper grain boundary dynamics during coupled motion and grain rotation to show the abilities and limitations of the amplitude equations description.

The article is organised as follows:
First, in section \ref{model::section} we introduce the amplitude equations model and discuss its relation to classical phase field modelling and the density functional theory of freezing.
In particular we show that the fact that many elements in the periodic table crystallize in a body centred cubic (bcc) structure first when solidified from the melt phase, is reflected also in this model.
In section \ref{PBC_CSL} we discuss the role of periodic boundary conditions, as they are frequently used for spectral implementations of the model.
Here we discuss in detail how the constraints on the system size in order to fulfil all periodicity conditions are related to coincidence site lattices (CSLs).
Section \ref{coupled::section} is devoted to the coupling dynamics, as modelled by the amplitude equations.
Here we consider two scenarios which are important for many metallurgical applications, namely the coupled motion of grain boundaries, which are subjected to a shear force, and grain rotation.
In section \ref{nonlinear} we discuss nonlinear elastic deformations, and how they are represented in the amplitude equations model.
Finally, the results are summarised in section \ref{summary}.

\section{Model description}
\label{model::section}

In this section we introduce the amplitude equation model, which is also called Ginzburg-Landau model.
It can be derived rigorously via a multiscale expansion from the phase field crystal model, and it can also be linked to classical density functional theory, from which it can be obtained using the Ramakrishnan-Yussouff functional \cite{Emmerich:2012uq}.
We refrain here from a derivation of the model and instead refer to \cite{Wu:2006uq,Spatschek:2010fk}.

Conceptually, the amplitude equations model can be understood as an approach for a phase field model with atomic resolution.
For that, let us briefly recapitulate the basics of a phase field model \cite{ElderBook,ChenReview,BoettingerReview,KarmaReview,steinbach,Spatschek11}.
In the simplest case, a single order parameter $\phi(\vec{r}, t)$ is introduced, which has specific values inside a phase.
As an example, we can use $\phi=0$ for a liquid and $\phi=1$ for a solid phase.
At the interface between them, one uses a smooth interpolation between these two bulk states on a length scale $\xi$, which is a numerical parameter.
In contrast to a sharp interface theory, where the positions of the interface are tracked explicitly in a dynamical simulation of e.g.~solidification or melting, the motion of the interfaces is expressed via an evolution equation for the order parameter.
Often, one uses variational formulations based on a free energy functional $F$, such that the equation of motion has the structure
\begin{equation}
\frac{\partial \phi}{\partial t} = -K\frac{\delta F}{\delta \phi}.
\end{equation}
Here, $K$ is a kinetic coefficient.
There are two central points, which we will discuss in particular in comparison with the amplitude equations model:
(i) The order parameters are spatially constant within each phase, therefore the model does not resolve any substructure. 
(ii) Coexistence of the phases demands that the free energy landscape $F$ has two minima with equal value for the bulk states at the coexistence temperature.
This is often realised by the use of a double well potential in the free energy functional.
In its simplest form, the free energy is
\begin{equation}
F = \int d\vec{r} \left[ \frac{a}{2} \phi^2(1-\phi)^2 + \frac{b}{2} (\nabla \phi)^2 + L \frac{T-T_M}{T_M} h(\phi) \right],
\end{equation}
where the first term proportional to $\phi^2(1-\phi)^2$ is the aforementioned double well potential with minima at $\phi=0$ and $\phi=1$, and the gradient square term penalises sharp phase field gradients.
The last term, which involves the latent heat $L$, favours either the solid or melt, depending on the deviation of the temperature $T$ from the melting point $T_M$.
Moreover, $a$ and $b$ are constants which determine the interface thickness $\xi\sim(b/a)^{1/2}$ and $h(\phi)$ is a switching function with the property $h(0)=0$ and $h(1)=1$.

\begin{figure}
\begin{center}
\includegraphics[width=6cm]{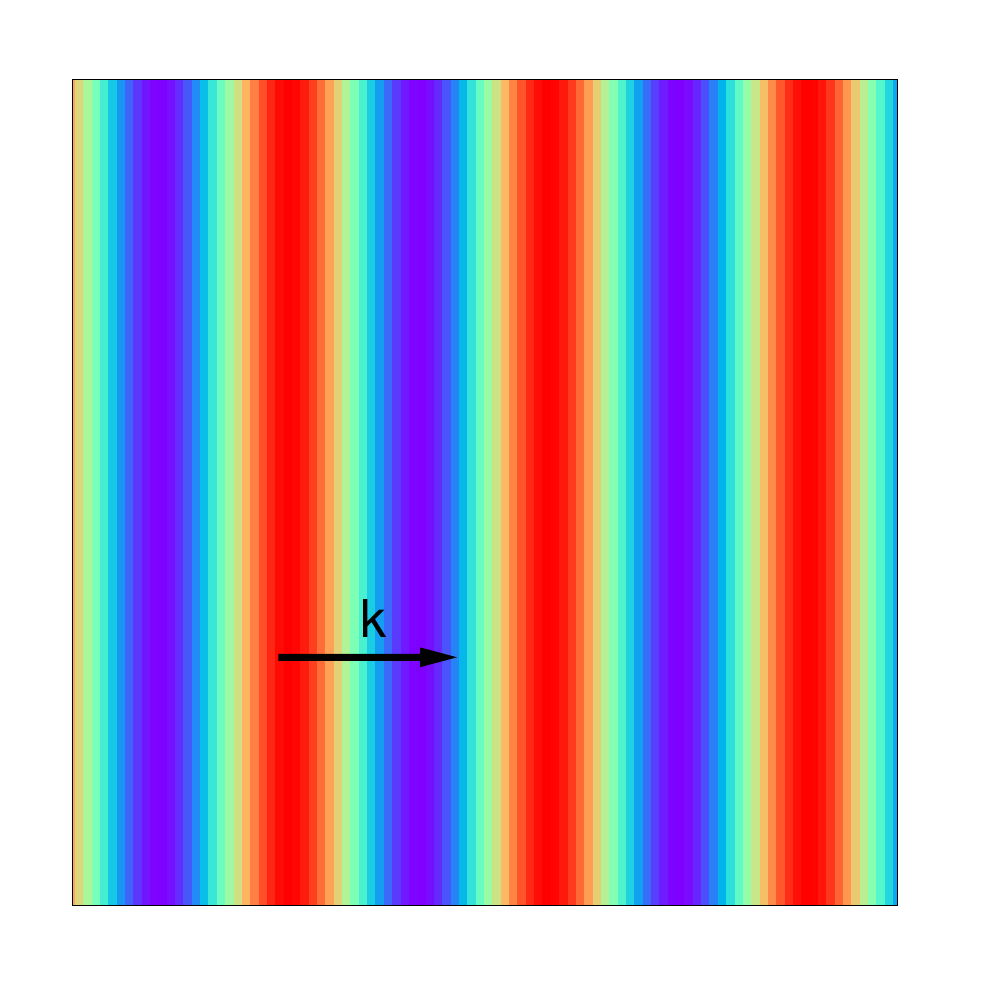}
\includegraphics[width=6cm]{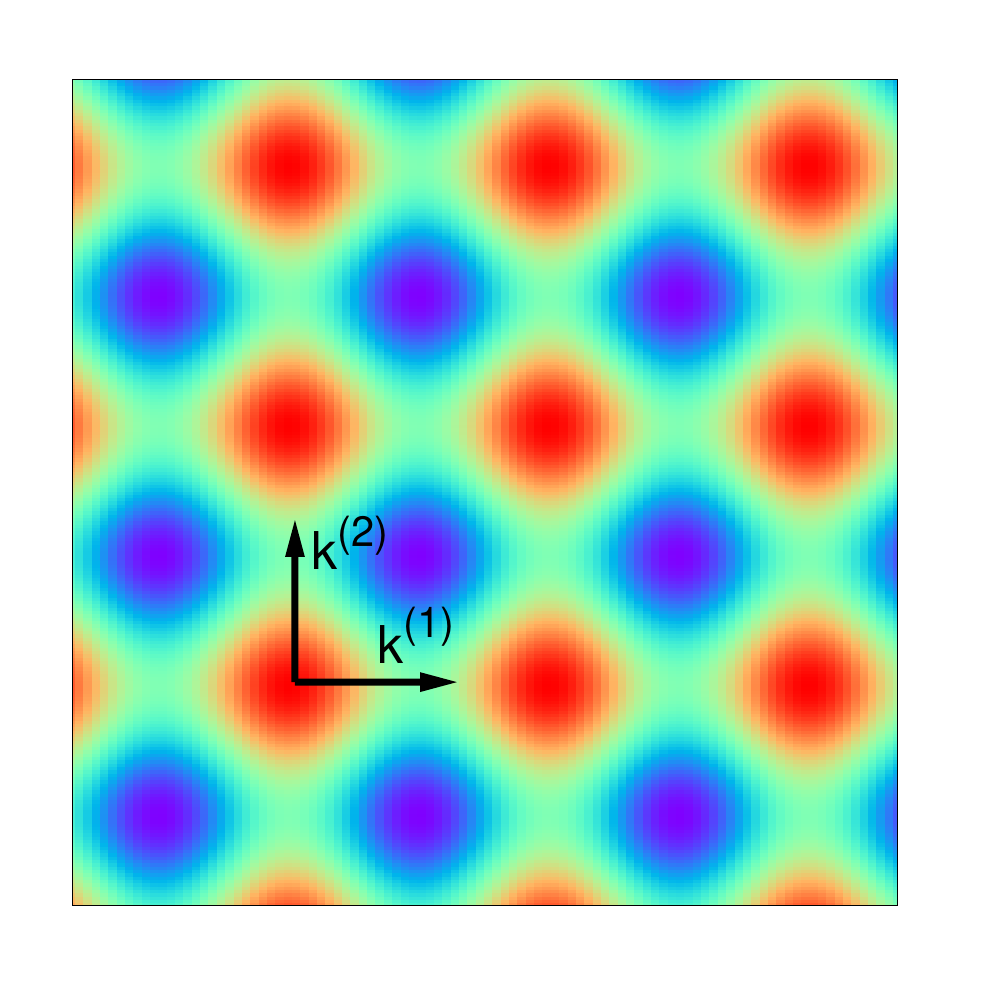}
\includegraphics[width=6cm]{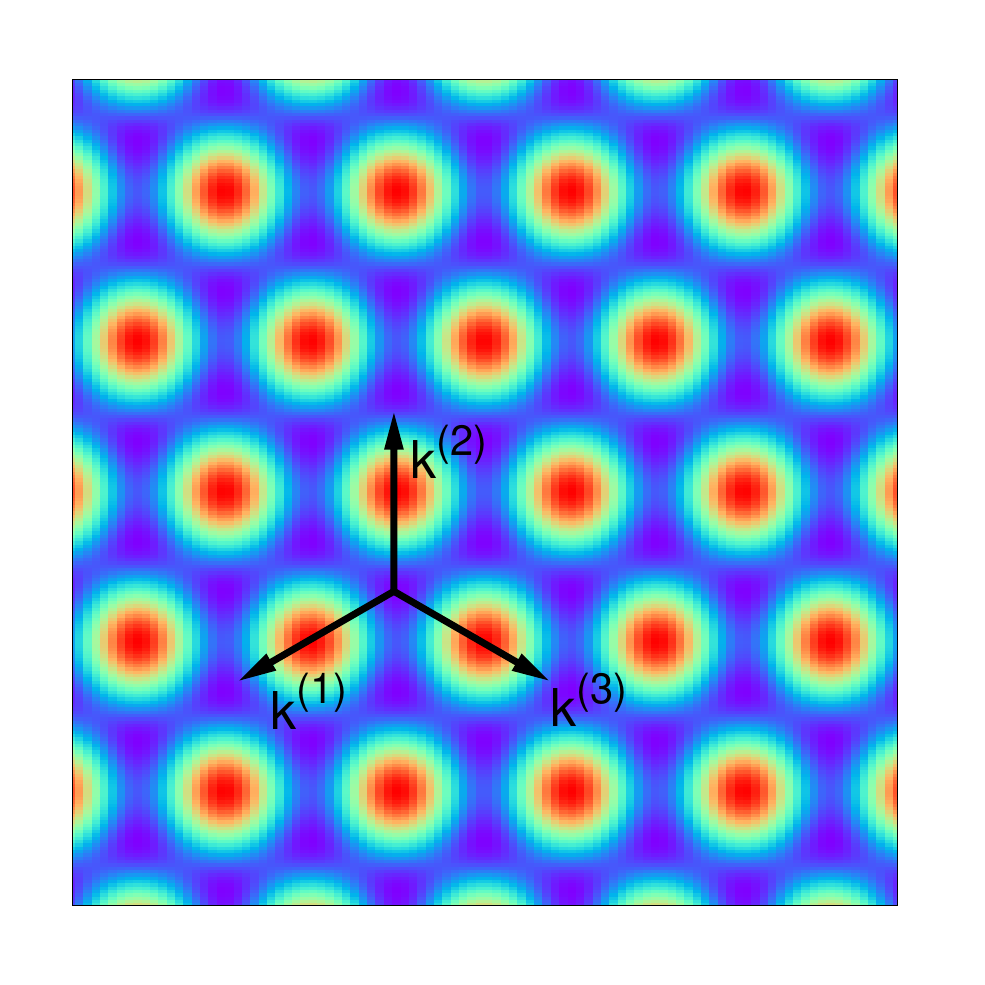}
\includegraphics[width=6cm]{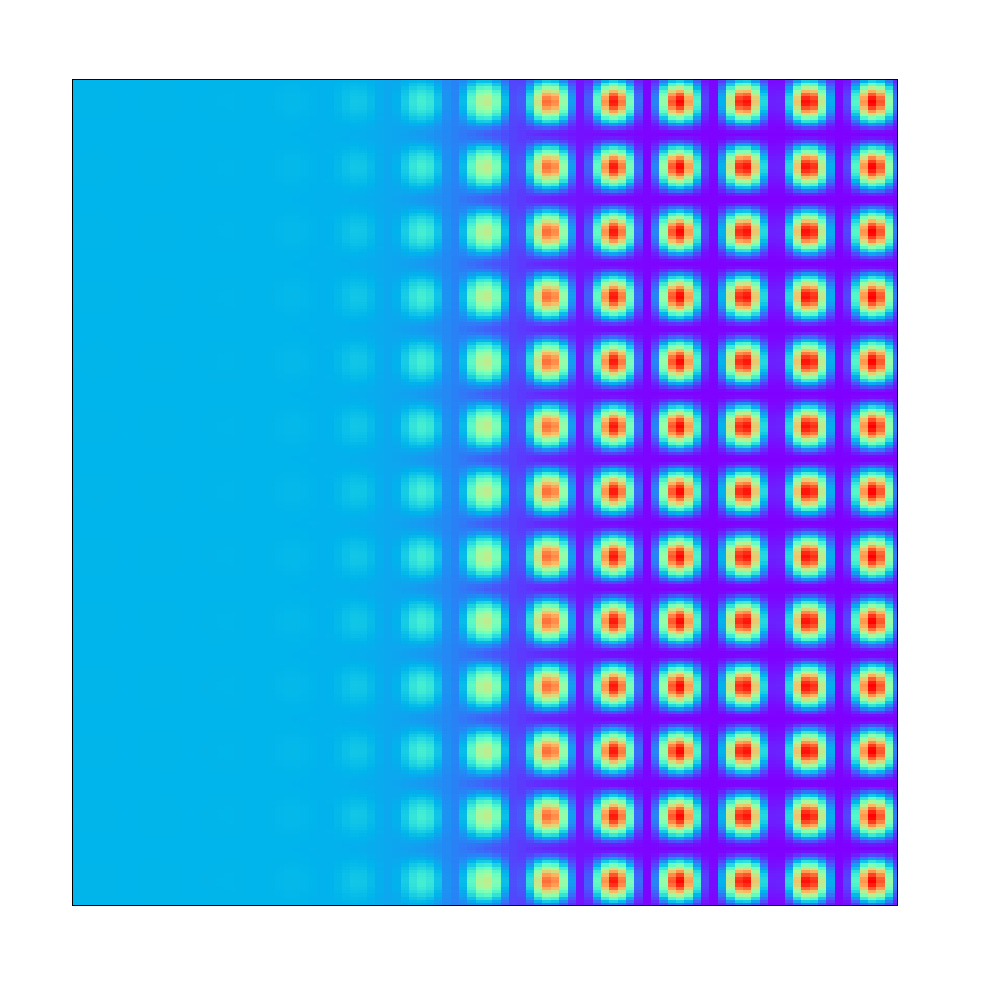}
\caption{Reconstructed atomic densities for constant amplitudes $u^{(j)}$ and different sets of reciprocal lattice vectors (RLVs) in two and three dimensions.
(a) Only one RLV (plus the inverse vector), $\vec{k}=(1,0)$ leads to the smectic phase.
(b) Two perpendicular vectors $\vec{k}^{(1)}=(1,0)$ and $\vec{k}^{(2)}=(0,1)$ (plus inverse vectors) represent a simple cubic lattice in two dimensions.
(c) With three vectors $\vec{k}^{(1)}=(-\sqrt{3}, -1)/2$, $\vec{k}^{(2)}=(0, 1)$ and $\vec{k}^{(3)}=(\sqrt{3}, -1)/2$ (+ inverse vectors) a two-dimensional hexagonal lattice is formed.
(d) With the set of 6+6 RLVs as given in the text a body centred cubic (bcc) crystal is described.
Here it is shown as a planar solid-melt interface, where the amplitudes are zero in the left half and have a finite and equal value in the right half of the system.
}
\label{fig1}
\end{center}
\end{figure}

In contrast, for the amplitude equations model we assume that the atomic density is given by
\begin{equation}
n(\vec{r})= n_0 + \sum_j \uj e^{i\vec{k}^{(j)}\cdot\vec{r}},
\label{cexp}
\end{equation}
with a background density $n_0$.
In its simplest form, as it is derived from the phase field crystal model \cite{Wu:2007kx,Spatschek:2010fk}, we do not take into account a density difference between solid and liquid.
The summation runs over a set of {\em principal reciprocal lattice vectors} $\vec{k}^{(j)}$, which will be specified below.
Each term corresponds to a plane wave contribution, which is weighted by an amplitude $u^{(j)}$, which is also position dependent.
In a melt phase, where the atoms move freely, the (time averaged) atomic density is spatially constant, and therefore all amplitudes vanish, $u^{(j)}=0$.
In contrast, in a solid phase, they have a finite value, and then the density has a periodic modulation, which corresponds to a certain lattice structure. 
At the positions of the atoms the density is high and low in between.
Examples for such structures are visualised in Fig.~\ref{fig1}.
Since the atomic density is a real quantity, the set of reciprocal lattice vectors (RLVs) contains pairs of antiparallel vectors, $\vec{k}^{(j)}+\vec{k}^{(\bar{j})}=0$, and the corresponding amplitudes are complex conjugate, $u^{(j)}=u^{(\bar{j})*}$.
As an example, for body centred cubic (bcc) crystals, which will be in the focus of this article, the set of RLVs consists of $[110], [101], [011], [1\bar{1}0], [10\bar{1}], [01\bar{1}] $ and their inverses $[\bar{1}\bar{1}0], [\bar{1}0\bar{1}], [0\bar{1}\bar{1}], [\bar{1}10], [\bar{1}01], [0\bar{1}1]$.
In sum, these are $N=12$ vectors and amplitudes, but only 6 of them are independent.

The analogous expressions to the order parameter $\phi$ in a phase field model are the amplitudes $u^{(j)}$.
The idea is that they vary on length scales which are large in comparison to the scale of the atomic oscillations $1/q_0=1/|\vec{k}^{(j)}|$.
A first obvious difference to a classical phase field model is that more than one order parameter is needed even to describe just a single solid phase.
Second, the order parameters are complex, and their phase encodes elastic deformations, lattice shifts and rotations.
We will come back to this point later.

Before we give an explicit expression for the free energy, from which the equations of motion are derived in particular for the applications in this article, we first discuss a more general expression, as it is obtained from density functional theory.
Several approximations lead to the following generic free energy \cite{Wu:2006uq},
\begin{eqnarray}
F &=& \frac{n_0 \kb T}{2} \int d\rv \Bigg[ \frac{1}{S(q_0)} \sum_{j=1}^N |\uj|^2 - \frac{C''(q_0)}{2} \sum_{j=1}^N |\Box_j \uj|^2 \label{deltakeq} \\
&&- a_3 \sum_{ijk}\alpha_{ijk} u^{(i)} u^{(j)} u^{(k)} \delta_{0, \vec{k}^{(i)}+\vec{k}^{(j)}+\vec{k}^{(k)} } + a_4 \sum_{ijkl} \alpha_{ijkl} u^{(i)} u^{(j)} u^{(k)} u^{(l)}\delta_{0, \vec{k}^{(i)}+\vec{k}^{(j)}+\vec{k}^{(k)}+\vec{k}^{(l)} } \Bigg],\nonumber 
\end{eqnarray}
which is valid independent of the underlying crystal structure.
The expression involves the Boltzmann constant $\kb$ and the Fourier transform of the direct correlation function $C(q)$, which is evaluated at the first peak $q_0$ of the structure factor $S(q)=[1-C(q)]^{-1}$.
The structure factor and the direct correlation function can be determined in molecular dynamics simulations or via scattering experiments \cite{Wu:2006uq}.
The operator $\Box_j$ is a generalisation of the gradient square term in a phase field model and reads \cite{PhysRevLett.80.3888,PhysRevE.50.2802,PhysRevLett.76.2185,Spatschek:2010fk}
\begin{equation} \label{boxdef1}
\Box_j=\kh_j\cdot\nabla - \frac{i}{2 q_0} \nabla^2
\end{equation}
with the normalised RLVs $\kh_j=\kv_j/|\kv_j|$.
The coefficients $a_i$, $\alpha_{ijk}$ and $\alpha_{ijkl}$ depend on the crystal structure and the underlying model.
For an equal weight ansatz for bcc crystals one obtains for example \cite{Wu:2006uq}
\begin{equation}
\alpha_{ijk}=1/8,\quad \alpha_{ijkl}=1/27,\quad a_3=\frac{24}{S(q_0)u_s},\quad a_4=\frac{12}{S(q_0)u_s^2}, 
\end{equation}
where $|\uj|=u_s$ is the constant amplitude in the solid phase.
We note that these coefficients differ slightly from the ones obtained from the PFC model \cite{Wu:2007kx}, which will be used as starting point below.
However, the difference appears only in the quartic term $\alpha_{ijkl}$ and has only a tiny influence e.g.~on the solid-melt interface anisotropy.

A central feature of the above free energy expression is that it gives in the local terms only contributions if the actual set of involved RLVs form a closed polygon, as expressed through the $\delta$-function, which is one if the two subscripts coincide and otherwise zero.
Notice that also the quadratic term proportional to $S(q_0)^{-1}$ has this structure, where the factor $\delta_{0,\vec{k}^{(i)} +  \vec{k}^{(j)} }$ has been used to reduce a double sum to a single one by the observation that only the combination $\vec{k}^{(i)} + \vec{k}^{(\bar{i})}$ with the inverse vector (and corresponding complex conjugate amplitude $\ujs$) can lead to a non-vanishing contribution.
In a solid phase all amplitudes belonging to the principal set of RLVs appear with equal magnitude $u_s$ due to the symmetry of the crystal.
Altogether, the terms which are quadratic, cubic and quartic in the amplitudes $u_j=u_s$ for a crystal which is neither deformed nor rotated, correspond to the terms in the classical phase field double well potentials $\phi^2(1-\phi)^2=\phi^2 - 2\phi^3 + \phi^4$.
As we have pointed out before, the existence of a second minimum apart from the trivial one $\phi=0$ is essential for having phase coexistence between solid and liquid.
Apparently, the negative cubic term $-2\phi^3$ plays here a central role, as without it the potential would have only one equilibrium state $\phi=0$.
It is now instructive to consider different crystal structures.
For bcc, the set of principal RLVs belongs to the face centred cubic (fcc) lattice, which are given above.
As one can readily check, they allow to form closed triangles of RLVs.
An example for this is $[0\bar{1}\bar{1}]+[101]+[\bar{1}10]=0$, and therefore the model indeed has a term which is cubic in the amplitudes.
According to the above discussion, coexistence between solid and melt is therefore possible.
In contrast, to describe an fcc crystal, one would use here as principal RLVs the bcc lattice vectors $\langle111\rangle$.
However, with them it is not possible to form a closed triangle, hence in such a description solid-melt coexistence would not be possible.
We mention that the observation, that many elements solidify first in bcc (and only at lower temperatures convert to the more densely packed structures like fcc) is therefore in line with the amplitude equations model.
For a more involved discussion of this issue we refer to \cite{Alexander:1978kx}.
In turn, this limitation implies that modelling of fcc structures requires to include additional RLVs, as done in \cite{Wu:2010vn}. 

To be more explicit, we use in the following the bcc amplitude equations description as derived from the PFC model \cite{Elder:2004ys,PhysRevLett.88.245701,ElderBook,Wu:2007kx,RevModPhys.65.851}.
A detailed derivation is given in \cite{Wu:2006uq,Spatschek:2010fk}.
The free energy functional reads
\begin{eqnarray}
F_{AE} &=& F_0 \int d\Rv \Bigg[ \sum_{i=1}^{N/2} |\Box_j \Aj |^2 + \frac{1}{12} \sum_{j=1}^{N/2} \Aj\Ajs 
+ \frac{1}{90} \Bigg\{ \left( \sum_{j=1}^{N/2} \Aj\Ajs \right)^2  - \frac{1}{2}\sum_{j=1}^{N/2} |\Aj|^4 \nonumber \\
&&+ 2A_{110}^* A_{1\bar{1}0}^* A_{101} A_{10\bar{1}} + 2A_{110} A_{1\bar{1}0} A_{101}^* A_{10\bar{1}}^* 
+ 2A_{1\bar{1}0} A_{011} A_{01\bar{1}} A_{110}^* 
\nonumber \\
&& 
+ 2A_{1\bar{1}0}^* A_{011}^* A_{01\bar{1}}^* A_{110} 
+ 2A_{01\bar{1}} A_{10\bar{1}}^* A_{101} A_{011}^* + 2A_{01\bar{1}}^* A_{10\bar{1}} A_{101}^* A_{011} \Bigg\} 
\nonumber \\
&&
-\frac{1}{8} \big( A_{011}^* A_{101} A_{1\bar{1}0}^* + A_{011} A_{101}^* A_{1\bar{1}0} + A_{011}^* A_{110} A_{10\bar{1}}^* 
+ A_{011} A_{110}^* A_{10\bar{1}} + A_{01\bar{1}}^* A_{110} A_{101}^* \nonumber \\
&&
+ A_{01\bar{1}} A_{110}^* A_{101} 
+ A_{01\bar{1}}^* A_{10\bar{1}} A_{1\bar{1}0}^* + A_{01\bar{1}} A_{10\bar{1}}^* A_{1\bar{1}0} \big) \Bigg] + F_T. \label{fenAE}
\end{eqnarray}
Here we have introduced a dimensionless ``slow'' length scale
\begin{equation} \label{slow}
\Rv =\epsnew^{1/2}q_0 \rv
\end{equation}
with
\begin{equation} \label{gl::eq1}
\epsnew = - \frac{24}{S(q_0) C''(q_0) q_0^2}
\end{equation}
and rescaled amplitudes 
\begin{equation}
\Aj = \uj/u_s.
\end{equation}
The length unit $\Rv$ is the scale of the diffuse interface thickness, in contrast to the ``fast'' scale $\vec{r}$ of the atomic oscillations, and their separation is the basis for the underlying multiscale analysis.
The differential operator $\Box_j$ is on the slow dimensionless scale
\begin{equation} \label{box}
\Box_j = \khj\cdot \nabla - \frac{i\epsnew^{1/2}}{2} \nabla^2,
\end{equation}
where the nabla operator acts on the slow scale $\Rv$.
The common prefactor of the free energy functional is given by
\begin{equation} \label{F0}
F_0 = - \frac{n_0 k_B T_M}{2} C''(q_0) q_0^{-1} u_s^2 \epsnew^{-1/2}.
\end{equation}
The thermal tilt is 
\begin{equation} \label{tilt::eq4}
F_T = L \frac{T-T_M}{T_M} \epsnew^{-3/2} q_0^{-3} \int d\Rv  \sum_{j=1}^{N/2} \frac{2\sqrt{\uj \ujs}}{N u_s}.
\end{equation}
The choice of the coupling function is discussed in more detail in \cite{Adland:2013ys,Spatschek:2010fk}.

For bcc $\delta$-iron the parameters are explicitly \cite{Wu:2006uq,Spatschek:2010fk}:
$q_0=2.985\times10^{10}\,\mathrm{m}^{-1}$, $S(q_0)=3.01$ and $C''(q_0)=-10.4\times 10^{-20}\,\mathrm{m}^2$, hence $\epsnew=0.086$.
In comparison to the fast dimensionless scale $q_0\vec{r}$, which varies on the scale of the atomic distances $a\sim1/q_0$, the ``slow'' scale (\ref{slow}) changes on the scale of the solid-melt interface thickness $\xi\sim \epsnew^{-1/2} q_0^{-1}$.
The entire free energy functional is written on the slow scale, and the atomic oscillations can be reconstructed using relation (\ref{cexp}).

Thermodynamic equilibrium corresponds to a stationary state of the free-energy functional.
We use relaxation dynamics
\begin{equation} \label{aemain}
\frac{\partial \Aj}{\partial t} = - K_j \frac{\delta F}{\delta \Ajs},
\end{equation}
with kinetic coefficients $K_j$, which we choose all to be the same, $K_j=K$.

A deformation, rotation or translation of a crystal leads to a change of the amplitudes according to
\begin{equation} \label{deformation}
\uj\to\uj \exp[-i\vec{k}^{(j)}\cdot\vec{u}(\vec{r})]
\end{equation}
with the displacement field $\vec{u}$.
This follows directly from the fact that the atoms are displaced from position $\vec{r}$ to $\vec{r}+\vec{u}(\vec{r})$ and the comparison with the expression (\ref{cexp}).
In particular, a rigid body translation leads to a constant shift of the phase of the amplitudes.
If the solid is deformed, its energy increases in agreement with the linear theory of elasticity, and the elastic constants have been computed in \cite{Spatschek:2010fk}.
Nonlinear elastic effects will be considered in Section \ref{nonlinear}.
In contrast to conventional phase field models, where elastic effects have to be added on top \cite{Spatschek:2007vn}, they are here contained in the description automatically.

\section{Periodic boundary conditions and coincidence site lattices}
\label{PBC_CSL}

In view of the aim to develop materials with superior properties special attention is paid to grain boundaries.
Their properties depend significantly in particular on the misorientation between the grains, and the resulting material properties can differ strongly.
{\em Grain boundary engineering} is therefore the practice to generate microstructures with a high fractions of grain boundaries with desirable properties.
Many of these properties are associated with boundaries that have a relatively simple, low energy structure.
Geometrically, these low energy structures are often associated with coincidence site lattices (CSLs), which are related to special grain boundaries.

The general concept of the CSL is a superstructure which can be imagined by overlapping two rotated grains and defining the coincident sites in both grains.
This purely abstract superstructure has the advantage that it allows to suggest low-energy states of a grain boundary.
Consequently, the preferability of grain boundary planes which contain as many CSL points as possible leads to the creation of small secondary defects to adapt to such a grain boundary structure.
However, it should be pointed out that not only the misorientation between the grains matters, but also the boundary plane.
Therefore, such a concept has its main use for pure tilt or pure twist boundaries.

Although the O-lattice theory of Bollmann \cite{BollmannBasics1972} offers a more intuitive model of grain boundary structures due to the continuous description in which preferred dislocation sites are predicted, CSLs are convenient for numerical modelling, as for many approaches periodic systems are used, and therefore corresponding boundary conditions naturally appear.
In general, a higher angle grain boundary has a shorter periodicity, and therefore can be simulated in a smaller system.
This is particularly important for ab initio simulations, as there only typically up to ${\cal O}(10^2)$ atoms can be simulated.

A central element is the introduction of the sigma value $\Sigma$, which is the ratio of the size of a unit cell formed by the coincidence lattice sites, relative to the size of the standard unit cell. 
For cubic crystals, this number $\Sigma$ is always odd.
It is related to the misorientation $\theta$ between the grains at a symmetric grain boundary.
A way to obtain this value and to relate it to the misorientation is to consider Pythagorean triplets of integer numbers $\{a_1,b_1,\Sigma_\phi\}$ with the property $a_1^2 + b_1^2 = \Sigma_\phi^2$.
In a geometrical interpretation we write $a_1=\Sigma_\phi\sin\phi$ and $b_1=\Sigma_\phi\cos\phi$, where the angle $\phi$ in the associated right-angled triangle is half the misorientation in a symmetric tilt grain boundary, $\phi=\theta/2$.

Also for the amplitude equations the use of spectral methods is beneficial, and therefore also here domain sizes should be chosen such that periodicity conditions are met.
If a grain is rotated relative to the reference set of reciprocal lattice vectors, the amplitudes are no longer constant inside the grain (even in the absence of elastic deformations), but undergo spatial oscillations,
\begin{equation}
A_j(\vec{r}) = A_s \exp(ik^{(j)\dagger} \mathbf{M} \vec{r}),
\end{equation}
with a matrix $\mathbf{M}=\mathbf{O}-\mathbf{1}$.
Here, $\mathbf{O}$ is an orthogonal rotation matrix and $\mathbf{1}$ the unity matrix \cite{Spatschek:2010fk}.
We consider here only symmetric tilts and therefore only in-plane rotations.
In the third direction, which we denote here as $z$ direction, the amplitudes are therefore translational invariant, and this trivial direction does not have to be considered in the following.
For a system of size $X\times Y$ (we measure the length in units of the inverse length of the principal RLVs) the periodicity conditions can then be reduced to
\bea
\vec{k}^{(j)\dagger} \bold{M} \displaystyle \binom{X}{0} &=& 2 \pi n_j, \\ 
\vec{k}^{(j)\dagger} \bold{M} \displaystyle \binom{0}{Y} &=& 2 \pi m_j,
\eea
with integer numbers $n_j$ and $m_j$.
Here we have to keep in mind that all amplitudes need to fulfil such periodicity conditions simultaneously.
It turns out that in particular for the bcc crystals, which we consider here, not all six complex amplitudes are independent of each other in this respect, but due to the fact that the reciprocal lattice vectors can form closed polygons, also the corresponding integer numbers $n_j$ are related.
An example is $[0\bar{1}\bar{1}]+[101]+[\bar{1}10]=0$, and therefore also $-n_{011} + n_{101} - n_{1\bar{1}0}=0$.
In the end, only two of the numbers $n_j$ can be chosen independently, as discussed in detail in \cite{Spatschek:2010fk}.
From the periodicity conditions in $x$ direction one arrives at
\begin{equation} \label{bceq1}
\tan \frac{\phi}{2} = \frac{n_{101}}{n_{011}},
\end{equation}
where $\phi$ is the aforementioned rotation of the crystal lattice relative to the fixed set of RLVs.
The minimum periodicity length is then
\begin{equation}
X = - \frac{2\sqrt{2}\pi n_{011}}{\sin{\phi}},
\end{equation}
where obviously $n_{011}$ has to be negative, and therefore also $n_{101}<0$ from the preceding formula for $\phi>0$.
An analogous consideration for the $y$ direction leads to
\begin{equation} \label{bceq2}
\tan\frac{\phi}{2} = - \frac{m_{011}}{m_{101}}
\end{equation}
and 
\begin{equation}
Y = \frac{2\sqrt{2}\pi m_{101}}{\sin\phi}.
\end{equation}
For practical purposes it is often desirable to choose the system size, which can accommodate the roatated grain (or a symmetric tilt grain boundary), to be as small as possible.
A particular challenge are then low angle grain boundaries (small misorientation $\theta=2\phi$), which suggest to choose $n_{101}=-1$, according to Eq.~(\ref{bceq1}).
For higher angle grain boundaries, such a choice is not possible.
We identify now the integer number $a_1$ in the Pythagorean triplet with the ``quantisation'' of the system size according to $a_1=-n_{011}$.
Then we readily get
\begin{equation}
X = 2\sqrt{2}\pi \Sigma_\phi,
\end{equation}
which relates the minimum system size to the coincidence site lattice once we describe the relation between $\Sigma_\phi$ and $\Sigma$, which is the value for the grain boundary. 
Using trigonometric identities
and defining the triplet $\{ a_2, b_2, \Sigma\}$, such that $\cos \theta = b_2/\Sigma_{\theta}$ and $\sin \theta =  a_2/\Sigma_{\theta}$, we obtain
\bea 
\frac{b_2}{\Sigma} &=& 2 \left(\frac{b_1}{\Sigma_{\phi}}\right)^2 -1,\\
\frac{a_2}{\Sigma} &=&  2\left( \frac{b_1}{\Sigma_{\phi}}\right)\sqrt{1-\left( \frac{b_1}{\Sigma_{\phi}}\right)^2}.
\eea
This yields finally for a rotation of each half grain by $\phi = \theta/2$ the periodicity coefficients $n_i, m_j$ and the triplet that describes the grain boundary as 
$\{ a_2,b_2,\Sigma\} = \{ 2 a_1 b_1, b_1^2 - a_1^2,\Sigma_{\phi}^2 \} $.

Specific examples of lattice rotations or grain boundaries with corresponding numbers $n_j$ are listed in Table \ref{table1}. 
\begin{table}
\caption{Choices of symmetric tilt grain boundaries.}
\label{table1}
\begin{tabular}{c|c|c|c|c|c}
$\phi = \theta/2$ & $\Sigma_\phi$ & triplet $\{a,b,\Sigma_\phi\}$ & $n_{011}$ & $n_{101}$ & $\Sigma$ \\
\hline
$0.154 = 8.797^\circ$    & 85 & $\{13,84,85\}$ & -13 & -1 & 7225\\
$0.181 = 10.389^\circ$  & 61 & $\{11,60,61\}$ & -11 & -1 & 3721\\
$0.221 = 12.680^\circ$  & 41 & $\{9,40,41\}$  & -9 & -1 & 1681\\
$0.284 = 16.260^\circ$  & 25 & $\{7,24,25\}$  & -7 & -1 & 625\\  
$0.644 = 36.87^\circ$ & 5 & $\{3,4,5\}$ & -3 & -1 & 25\\
$0.761= 43.603^\circ$   & 29 & $\{20,21,29\}$ & -20 & -8 & 841\\
\hline
\end{tabular}
\end{table}
Apparently, the smallest CSL that can be simulated in this way, having the proper periodicity behavior is $\Sigma = 25$.
Here, for a $\theta = 16.26^\circ$, we find $\theta/2 = 8.13^\circ$ or via symmetry $(\theta-\pi/2)/2 = -36.87^\circ$, which is equivalent, as the tilts $\theta$ are symmetric under rotations by $\pi/2$. 
Larger tilt angles can be reached with pythagorean triplets that are not constructed as $\{\sqrt{\Sigma^2-(\Sigma-1)^2}, \Sigma-1,\Sigma\}$, see the example for $\theta = 87.206^\circ$ in the last row of table \ref{table1}. However, due to the limitation of the current bcc amplitude equation model to rotations $\theta < \pi/4$, it is required to study the equivalent tilt $\pi/2 - \theta$ to stay in the regime of proper dynamics.

\begin{figure}
\includegraphics[trim=6cm 13cm 8cm 4cm, clip=true, width=14cm]{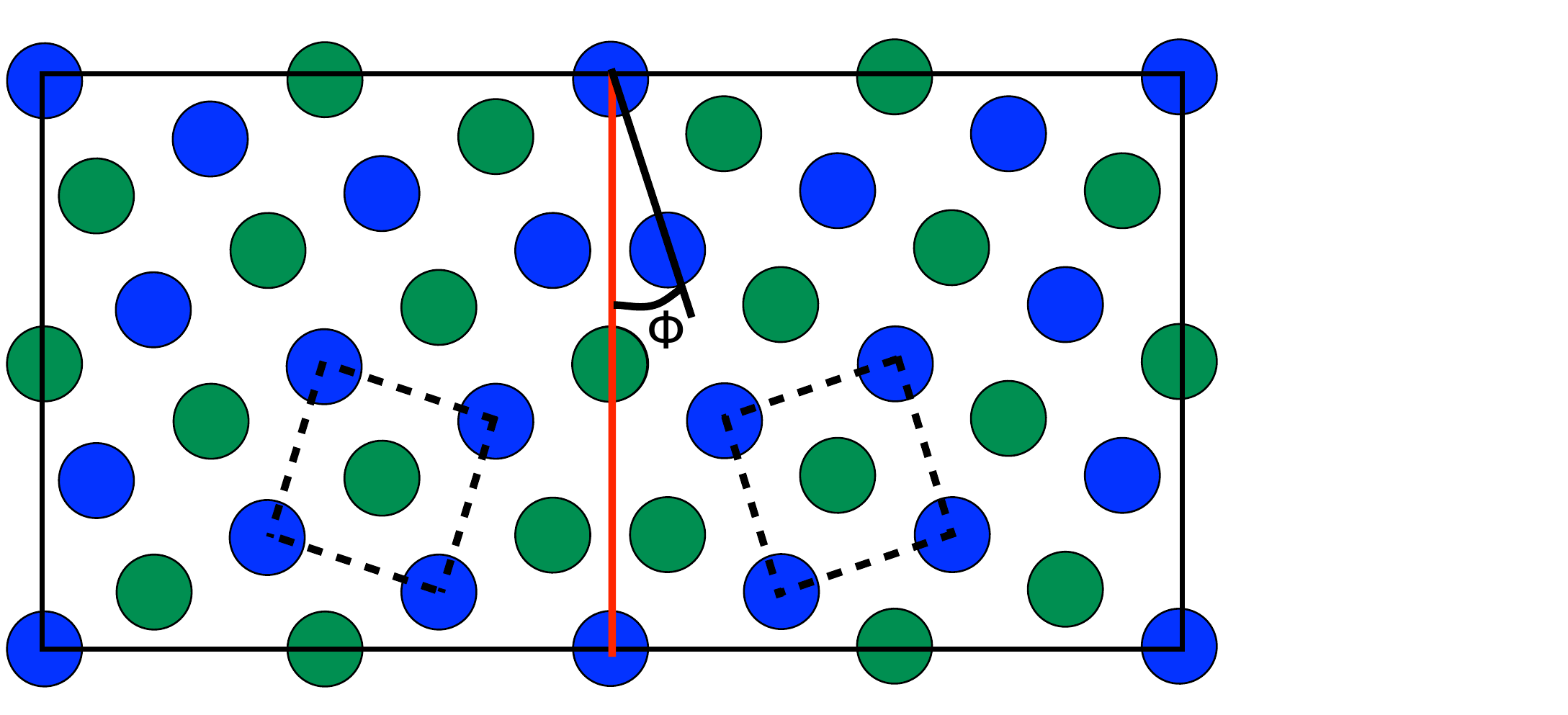}
\caption{$\Sigma 5\;(310)\;[001]\;36.87^\circ$ symmetric tilt grain boundary in bcc. 
The different colours of the atoms correspond to lattice planes which are shifted by half a lattice unit in out-of-plane direction.
The red line marks the grain boundary, and another is located at the left and right boundary, at which the supercell continues periodically.}
\label{tobedone}
\end{figure}

It should be pointed out that the periodicity conditions (\ref{bceq1}) and (\ref{bceq2}) are constraints for (i) the amplitudes and not for the (ii) atomic density.
As a consequence, periodicity of these entities (i) and (ii) is not equivalent.
A simple example to illustrate this difference is a case without rotation of the lattice.
Since then the amplitudes are spatially constant, there are {\em no} constraints on the periodicity, and consequently the system size can be chosen arbitrarily, despite the fact that atoms in the reconstructed density may be cut and non-periodic at the system boundary.
Another consequence of this difference is that symmetric tilt grain boundaries, which are easy to access e.g.~in ab initio simulations due to the small supercell size required for them, are not necessarily directly accessible in an amplitude equation simulation with periodic boundary conditions.
An example for this is a $\Sigma 5\;(310)\;[001]\;36.87^\circ$ symmetric tilt grain boundary, see also Fig.~\ref{tobedone}, which shows the supercell for a periodic atom density.
Each grain is rotated here by $\phi=\pm \arccos (3/\sqrt{10})$.
However, if we evaluate the condition (\ref{bceq1}) we obtain
\begin{equation}
\tan\frac{\phi}{2} = \frac{\tan\phi}{1+\sqrt{1+\tan^2\phi}} = \frac{1}{3+\sqrt{10}},
\end{equation}
which is not a rational number, and hence proper integer numbers $n_{101}$ and $n_{011}$ cannot be determined to satisfy Eqs.~(\ref{bceq1}) and similarly (\ref{bceq2}).
One can however always approximate such a boundary as close as desired, although this in general requires to simulate rather large systems.
An alternative is to use implementations without the need for periodic boundary conditions (e.g.~a real space code), or to embed the bicrystal in a liquid phase near the system boundary, such that periodicity conditions do not arise.
This however, may cause additional issues, as grain rotation may occur then.


\section{Coupled grain boundary dynamics}
\label{coupled::section}


\subsection{Coupled motion of planar grain boundaries}

\begin{figure}
\includegraphics[trim=0cm 9cm 0cm 0cm, clip=true, width=13cm]{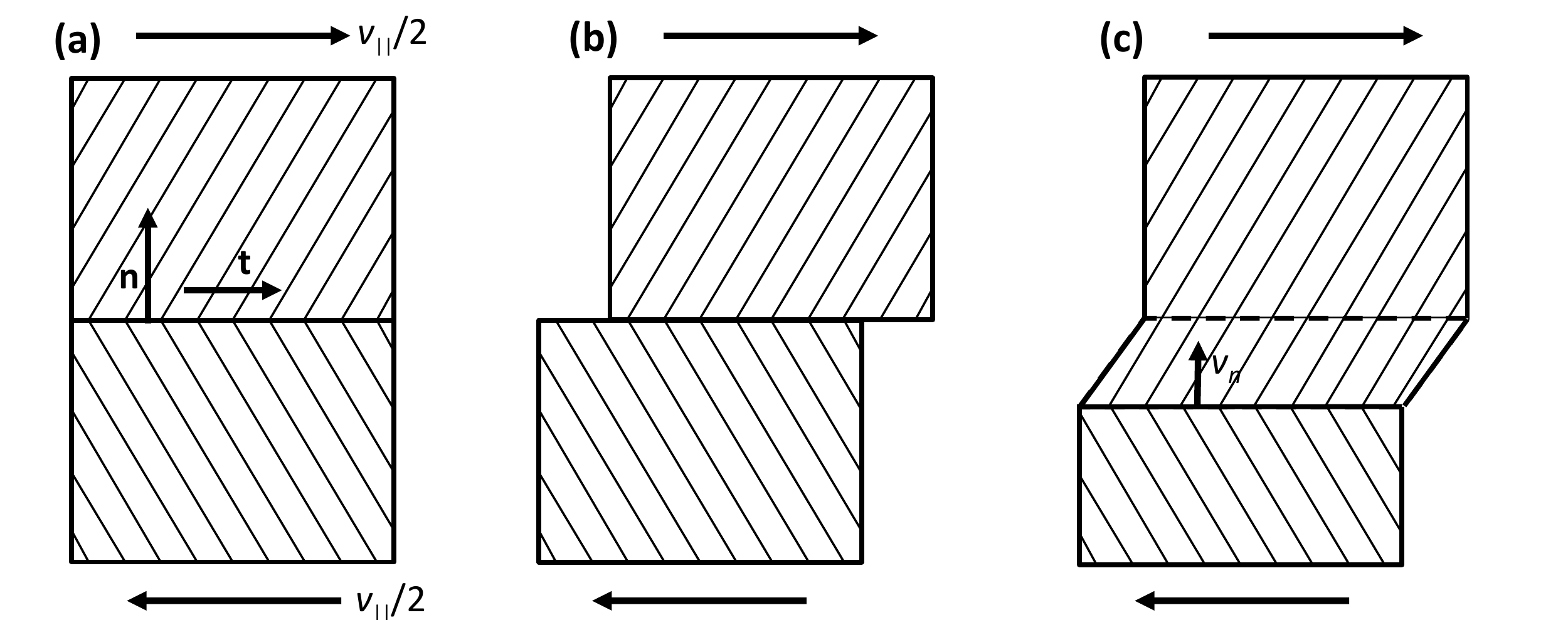}
\caption{Sketches of sliding and coupled motion of grain boundaries. 
(a) shows the original state of the crystal, which is subjected to tangential sliding, as indicated by the arrows. The shading illustrates the different grain orientations.
As a result of the shift, the two grains can either slide along each other, such that there is no normal motion of the grain boundary, as shown in (b).
During coupled motion, as shown in (c), atoms from one grain attach to the other one during the shear motion, and this effectively leads to a normal motion of the grain boundary.
}
\label{couplingsliding::fig1}
\end{figure}

Coupled and sliding motion of two crystals are related to grain boundary dynamics and appear, when two grains are sheared against other.
The geometrical situation is sketched in Fig.~\ref{couplingsliding::fig1}, where a tangential velocity difference between the crystals can cause a normal motion of the grain boundary in case of coupled motion.
Physically, atoms from one grain attach to the other grain while moving in the grain boundary plane.
As a result, a net motion of the grain boundary emerges.
Pure sliding motion, in contrast, implies a frictional motion of the grains without a shift of the grain boundary.
A comprehensive description in terms of a dislocation based perspective for low angle grain boundaries has been developed in \cite{ActaMater52_4887_2004_coupledGBMotion}.
These concepts have been confirmed by Molecular Dynamics simulations \cite{Cahn:2006fk}.
Phase field crystal simulations of transitions between coupled and sliding motion have been performed in \cite{Adland:2013ys}, where in particular a second order time derivative term has been used as proposed in \cite{PRE80_046107_2009_StefanovicHaatajaProvatas,arXiv_1306.5857_2013_GrasselliWu}, in order to separate the timescales for elastic relaxation and interface dynamics.
Simply speaking, this prevents e.g.~the unphysical bending of the lattice planes due to a elastic relaxation transported too slow relative to the shear rate. 

\begin{figure}
\includegraphics[width=\textwidth]{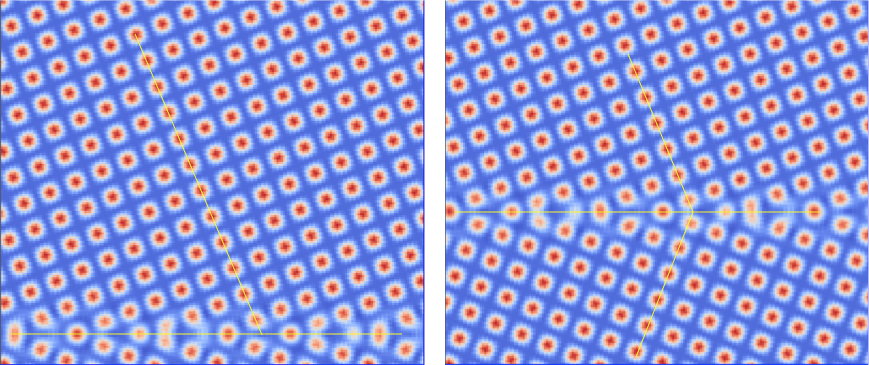}
\caption{Geometry of moving $\Sigma 169$ symmetric tilt grain boundary, as obtained from amplitude equations simulations. 
}
\label{geocoupledMotionGB} 
\end{figure}

Here we pursue the modelling of coupled grain boundary motion via the amplitude equation model.
To obtain quantitative information on the normal velocity of the grain boundary relative to the tangential velocity, we apply a displacement to one of the grains far away from the grain boundary, which induces a tangential motion, and the resulting normal velocity is measured.
We use a GPU implementation of the amplitude equations, as described in \cite{MSMSE_22_034001_2014_HueterNguyenSpatschekNeugebauer}, which benefits immensely from the spectral representation of the problem and allows to accelerate the code by two orders of magnitude in comparison to a single core CPU variant. 
The scheme to set up a certain grain boundary is described in section \ref{PBC_CSL}, and a close-up of the reconstructed density is depicted in Fig.~\ref{geocoupledMotionGB}. 
We define the normal direction of the interface $\vec{n}$ and chose the tangential direction $\vec{n}$ to be rotated by $\pi/2$ clockwise relative to $\vec{n}$, see Fig.~\ref{couplingsliding::fig1}. 
Accordingly, the normal growth direction $v_{n}$ is counted positive for motion in direction $\vec{n}$. 
The tangential velocity $v_{\parallel}$ is counted positive if the relative lateral motion of the grain is in direction $\vec{t}$.

As in \cite{ActaMater52_4887_2004_coupledGBMotion,Cahn:2006fk} the tangential motion of such a grain boundary is written as
\be
v_{\parallel} = S \sigma + \beta v_n,
\ee
with $\sigma$ being the tangential component of the applied stress, $S$ the sliding coefficient, and $\beta$ is denoted as coupling constant. 
The two limiting cases stated by this model are pure coupling, i.e. $v_{\parallel} = \beta v_n$, and pure sliding, which is described then as $v_{\parallel} = S \sigma$. 
Cahn and Taylor have worked out a geometrical model of coupling \cite{ActaMater52_4887_2004_coupledGBMotion}, which is based on an analysis of the dislocation distribution at low angle tilt boundaries.
For a symmetric tilt grain boundary no dislocation glide takes place unless $\sigma$ reaches the strength of the crystal.
Hence we expect $S=0$ in this case.
According to the geometrical model of coupled motion
\be
v_{\parallel}=2 \tan (\theta/2) v_n \approx \theta v_n,
\ee
which holds for low angle grain boundaries with $\langle100\rangle$ orientation.
Whereas the first relation is exact, the second holds for misorientations $\theta\ll1$.
In terms of the coupling constant therefore
\begin{equation}
\beta_{\langle100\rangle} = 2 \tan (\theta/2).
\end{equation}

For $\langle110\rangle$ orientation, the coupling is predicted to be 
\be
\beta_{\langle110\rangle} = - 2 \tan\left( \frac{\pi}{4} - \frac{\theta}{2} \right).
\ee

\begin{figure}
\includegraphics[width=0.85 \textwidth]{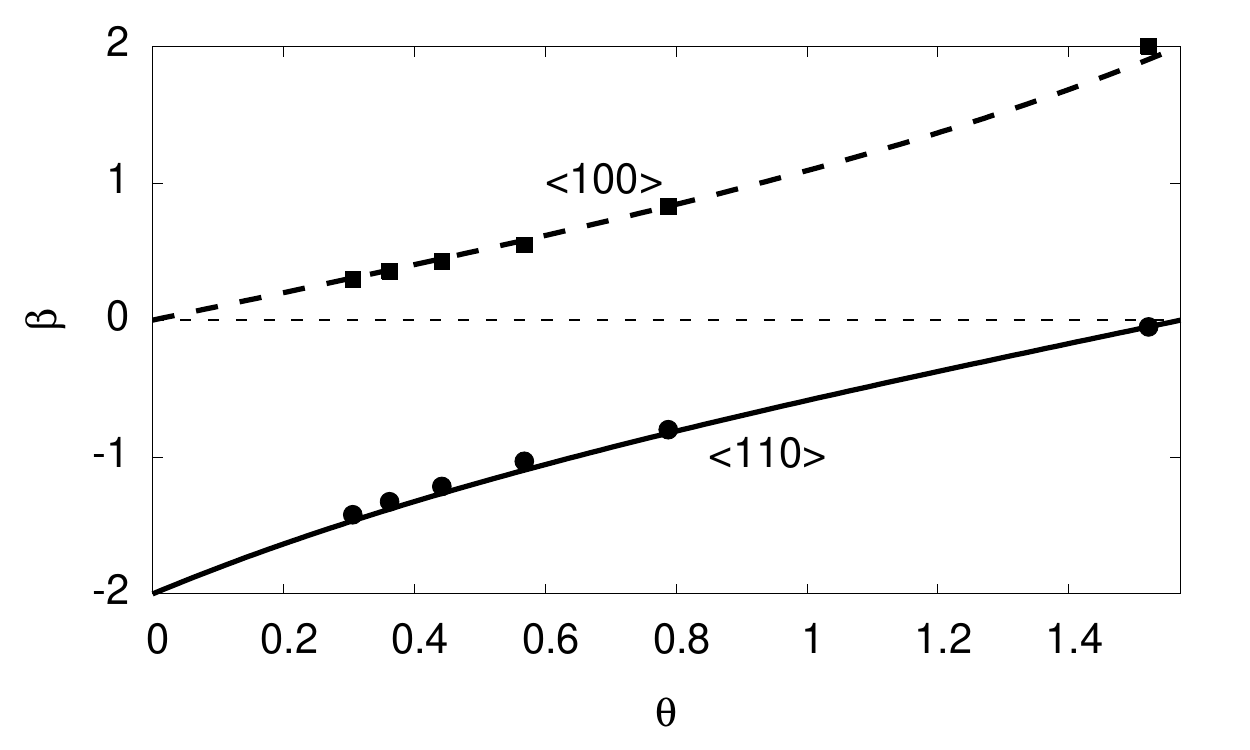}
\caption{Ratio of measured coupling coefficients $\beta_{\langle100\rangle}$ and $\beta_{\langle110\rangle}$ and the corresponding values predicted analytically as described in \cite{ActaMater52_4887_2004_coupledGBMotion}. The numerical results from the amplitude equations simulations (points) agree well with the theoretical prediction for pure coupled motion (curves).
}
\label{coupledGBResults}
\end{figure}

The results from the simulations are shown in Fig.~\ref{coupledGBResults} in comparison to the theoretical prediction.
Indeed, we find both coupling modes, and both of them show an excellent agreement with the theoretical prediction.
These results are obtained in the low temperature regime.
Additional simulations at high homologous temperatures show deviations from the coupling theory, as additionally sliding effects become visible, when due to premelting effects full coupling is no longer maintained.
This is in line with phase field crystal simulations in \cite{Adland:2013ys}, where a full phase diagram for the different coupling and sliding modes is extracted from the simulations.

There is however a central difference between the atomistic and phase field crystal simulations on the one hand and the amplitude equations descriptions on the other hand.
Whereas the first methods show a transition from $\langle 100\rangle$ to $\langle 110\rangle$ coupling if the misorientation is increased starting from a $(100)$ grain boundary, this does not occur for the amplitude equations.
The reason is related to the inability of the latter method to describe high angle grain boundaries correctly.
This has been discussed in \cite{Spatschek:2010fk} for the grain boundary energy $\gamma_{gb}$ as function of misorientation, where one would expect first a sharp increase of $\gamma_{gb}$ as function of the misorientation $\theta$ according to a Read-Shockley behavior.
Whereas this prediction is fully satisfied, the grain boundary energy does not decrease again if $\theta$ approaches $90^\circ$, where a perfectly healed crystal should form.
The amplitude equations however do not ``see'' this healing of the crystal, since due to the separation into the individual amplitudes an automatic change to a new ``reference set of RLVs'' does not happen.
This is an important limitation of the amplitude equations in their present form, and one should therefore keep in mind that they only deliver an accurate description for small rotation angles.
Here the same effect is reflected by the fact that the coupling mode cannot jump from the $\langle 100\rangle$ mode to $\langle 110\rangle$ branch, as they are not mutually accessible. 
Instead, one can follow both branches separately, provided that one starts the simulation from different reference RLV sets.


\subsection{Coupled motion of spherical grain boundaries and inclusions in grain boundaries}

\begin{figure}
\includegraphics[width=0.8 \textwidth]{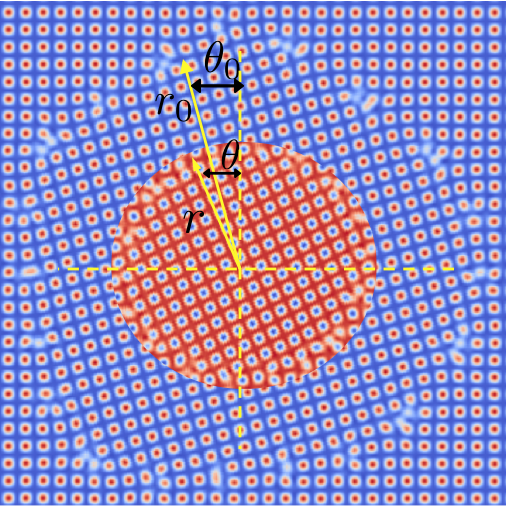}
\caption{Geometry of rotated inclusion. The deviations to a ideally circular shape are recognisable, but small in magnitude. 
The change from the large initial inclusion at $r_0, \theta_0$ to the shrunk, rotated inclusion at $r,\theta$ is shown by the inverted colour scheme for the rotated and shrunk inclusion.}
\label{geoRotInc}
\end{figure}

\begin{figure}
\includegraphics[width=0.8 \textwidth]{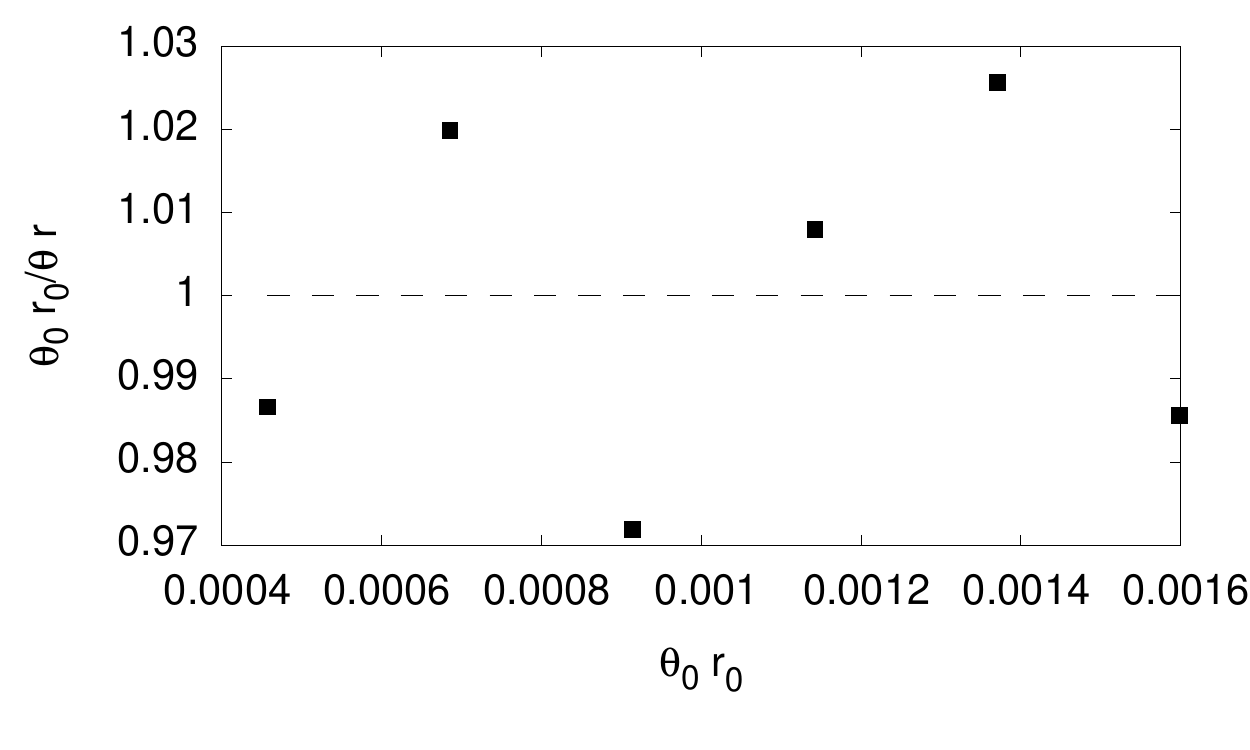}
\caption{Radial dependence of the rotation of the inclusion. 
The ratio $\theta r/ \theta_0 r_0$ is plotted versus the initial value of $\theta_0 r_0$, in agreement with the theoretical expectation (\ref{Cahn1}).
The simulations correspond to shrinking inclusions with the same misorientation and different radii $r_0$.
All data collapses to 1 as expected in Eq.~(\ref{Cahn1}).
}
\label{geothetavsradius}
\end{figure}

A spherical grain which is misoriented relative to its surrounding matrix can rotate during shrinkage
\cite{ActaMater52_4887_2004_coupledGBMotion}, which has also been observed in phase field crystal simulations \cite{Wu:2012fk}. 
We consider, as shown in Fig. \ref{geoRotInc}, a cylindrical crystal with a circular cross-section that is described by a radius $r(t)$ and misorientation $\theta(t)$. 
Apparently, rotation of the inclusion by $\Delta \theta$ corresponds to a relative displacement in tangential direction along the grain boundary, while radial movement means normal motion. 
Here we define $\vec{n}$ such that it points into the inclusion and the perpendicular tangential vector $\vec{t}$ is rotated counter-clockwise with respect to $\vec{n}$. 
The orientations of the velocities thus read
\bea
r(t) d\theta &=& v_{\parallel}dt, \\
dr &=& -v_{\vec{n}}dt.
\eea
Though an increase of the misorientation is energetically expensive, the overall energy is reduced as the interface shrinks, even when there is no bulk energy difference between the inclusion encircled by the curved grain boundary and the matrix phase. 
As derived in  \cite{ActaMater52_4887_2004_coupledGBMotion}, the equations describing the normal and tangential velocity are in the absence of a bulk energy difference
\bea
-v_{n} &=& \frac{dr}{dt} = -M \left( \frac{\gamma-\beta \gamma'}{r} + \beta \sigma \right), \label{claas1} \\
v_{\parallel} &=& r \frac{d \theta}{dt} = \beta M \left( \frac{\gamma - \beta \gamma'}{r} + \beta \sigma  \right) - S \left( \frac{\gamma'}{r} - \sigma \right). \label{claas2} 
\eea
Here, $\gamma(\theta)$ is the misorientation dependent grain boundary energy, $\beta$ the coupling constant, $S$ the sliding coefficient and $\sigma$ an externally applied stress.
In the following we focus on a situation without external stresses, $\sigma = 0$, and for pure coupling, $S=0$.
According to the system setup as shown in Fig.~\ref{geoRotInc} we measure the initial rotation angle and radius $\theta_0, r_0$, let the system evolve and measure $\theta=\theta_0+\Delta \theta, r= r_0-\Delta r$ afterwards.
For small angle tilt misorientations and $\langle100\rangle$ coupling, $\beta(\theta) = 2\tan(\theta/2)$. Consequently, 
we obtain by combining (\ref{claas1}) and (\ref{claas2})
\[
- \frac{dr}{dt} 2 \tan \frac{\theta}{2} = r \frac{d\theta}{dt},
\] 
and therefore get the relation $r \sin (\theta/2) = r_0 \sin (\theta_0/2)$. 
Here we further approximate for $\theta\ll1$ to obtain 
\be \label{Cahn1}
\frac{\theta r}{\theta_0 r_0} \approx 1.
\ee
This analytical expectation is well confirmed by the amplitude equations simulations, as shown in Fig.~\ref{geothetavsradius}.
%

\begin{figure}
\includegraphics[width=12cm]{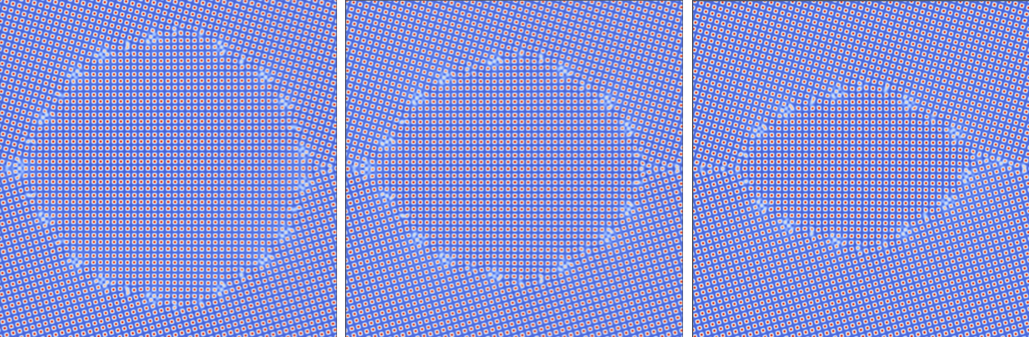}
\caption{Time evolution of the shrinkage of an initially spherical grain which is located on a $\Sigma 25$ symmetric tilt grain boundary.
Driven by the reduction of the interfacial energy the grain shrinks and acquires a lenticular shape.
Since equal portions of the inclusion are located below and above the grain boundary, the inclusion does not rotate due to a cancellation of the torques.
This effect in visualised more clearly in Fig.~\ref{blocked2}.
}
\label{blocked}
\end{figure}

\begin{figure}
\includegraphics[width=10cm]{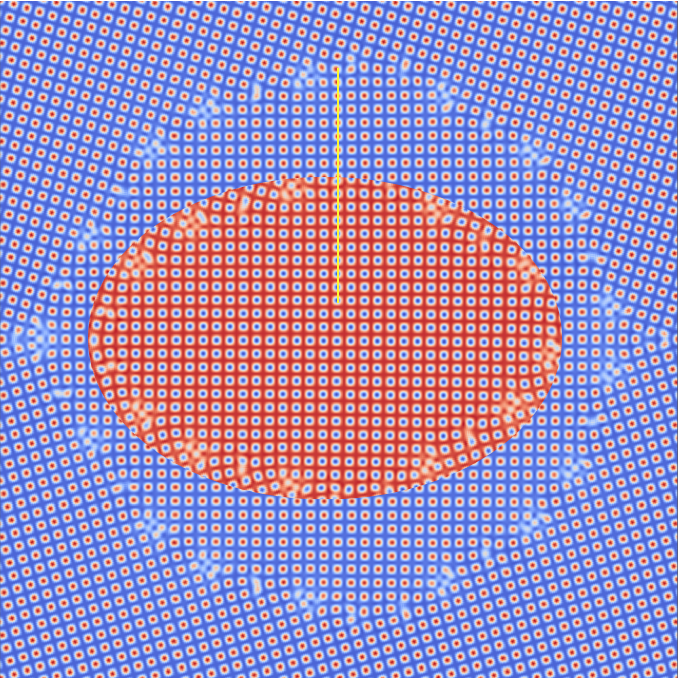}
\caption{Blocked rotation of an inclusion in a symmetric tilt GB. 
In contrast to Fig.~\ref{geoRotInc} the inclusion does not rotate here.
The figure shows the magnified initial and final states in Fig.~\ref{blocked}.
The inverted colour scheme of the shrunk inclusion at a later time shows that the lattice inside has not rotated.
}
\label{blocked2}
\end{figure}

The rotation of the grain is driven by the change of the interfacial energy as a combination of radius reduction and change of the misorientation.
If the misorientation $\theta$ is inverted, $\theta\to-\theta$, also the grain rotates in the opposite direction.
Consider a spherical grain located symmetrically on a straight symmetric tilt grain boundary. 
Assume that each half of this grain has exactly the opposite misorientation with respect to the  two half-crystals (as shown in Figs.~\ref{blocked} and \ref{blocked2}).
In this case, the torques exactly balance each other.
%
As a consequence, such a cylindrical inclusion does not rotate, and this is also reflected in amplitude equations simulations, in agreement with earlier phase field crystal simulations \cite{Wu:2012fk}.


\section{Rotational invariance and nonlinear elastic deformations}
\label{nonlinear}

It is an important benefit of the amplitude equations model that it automatically contains linear elasticity, and this has been investigated in detail in \cite{Spatschek:2010fk}.
The basic idea is that a deformation of the lattice, as described by Eq.~(\ref{deformation}), changes the energy density of a solid phase.
Here it is important that only the gradient contribution to the free energy changes, whereas the local terms remain the same.
The reason is that according to Eq.~(\ref{deltakeq}) the local terms only contribute if the involved RLVs form a closed polygon, as expressed through the $\delta$-functions.
Hence a factor like $u^{(i)}u^{(j)}u^{(k)} \sim \exp[-i(\vec{k}^{(i)}+\vec{k}^{(j)}+\vec{k}^{(k)})\cdot \vec{u}(\vec{r})]$ is independent of the displacement field $\vec{u}$, since the sum of the $\vec{k}$-vectors is always zero.
(We have to distinguish in the notation between the components $u_i$ of the displacement vector $\vec{u}$ and the amplitudes $u^{(i)}$.
For the latter superscripts are used.)
As has been shown in \cite{Spatschek:2010fk}, the energy increase is in the small strain regime given by 
\begin{equation} \label{elen}
f_{el} = \frac{1}{2} C_{ijkl}\epsilon_{ij}^\mathrm{lin}\epsilon_{kl}^\mathrm{lin}
\end{equation}
with the linearised elastic strain tensor 
\begin{equation} \label{linstrain}
\epsilon_{ij}^\mathrm{lin} = \frac{1}{2} \left( \partial_i u_j + \partial_j u_i \right).
\end{equation}
This allows to identify the elastic constants $C_{ijkl}$ which depend on the underlying crystal structure.
Per amplitude they have a contribution
\begin{equation}
C_{ijkl} = -\frac{n_0 \kb T}{2} \frac{C''(q_0)}{q_0^2} k_i k_j k_k k_l u_s^2,
\end{equation}
see \cite{Spatschek:2010fk} for details.
Altogether, for a 2D lattice with hexagonal symmetry the material becomes elastically isotropic, in agreement with the usual theory of elasticity.
For a 3D bcc structure, the elastic constants have been derived, and the material has a cubic symmetry also from point of view of linear elasticity.
What is important here is that from the entire nonlocal term proportional to $|\Box u|^2$ only the leading term involving $\vec{k}\cdot\nabla$ has been taken into account.
This is in the spirit of a small and long wavelength elastic deformation, where the second term, which contains $\nabla^2 u$, is negligible.

In the following we will investigate in more detail the role of this higher order derivative term.
To simplify the notation we consider only a single amplitude, noting that the entire elastic energy is the sum of the contributions from the individual modes, as stated in Eq.~(\ref{deltakeq}).
For a pure solid phase the amplitude reads then in agreement with Eq.~(\ref{deformation})
\begin{equation}
u = u_s \exp(-i\vec{k}\cdot\vec{u}),
\end{equation}
and one readily gets for the gradient term in the energy density
\begin{equation} \label{eq5}
|\Box u|^2 = u_s^2 q_0^{-2} \left[ k_\alpha k_\beta \partial_\alpha u_\beta + \frac{1}{2} k_\beta k_\alpha (\partial_\gamma u_\beta)(\partial_\gamma u_\alpha) \right]^2 + \frac{1}{4} u_s^2 q_0^{-2} \left[ k_\beta \partial_\alpha^2 u_\beta \right]^2
\end{equation}
according to the definition of the box operator given in Eq.~(\ref{boxdef1}).
The second ``bending'' term is negligible in the long wave limit, as it is assumed that the displacements change on a scale much larger than $1/q_0$.
The first term looks similar to the nonlinear strain tensor, but in fact it is different.
We define 
\begin{equation}
\bar{\epsilon}_{ij} = \frac{1}{2}  \left( \frac{\partial u_i}{\partial x_j} +  \frac{\partial u_j}{\partial x_i} +  \frac{\partial u_i}{\partial x_k}  \frac{\partial u_j}{\partial x_k}\right). \label{wrongtensor}
\end{equation}
Using this definition, we can rewrite the first term in the expression (\ref{eq5}) and obtain
\begin{equation}
|\Box u|^2 = u_s^2 q_0^{-2} (k_\alpha k_\beta \bar{\epsilon}_{\alpha\beta})^2+ \frac{1}{4} u_s^2 q_0^{-2} \left[ k_\beta \partial_\alpha^2 u_\beta \right]^2,
\end{equation}
which reminds of the structure of the elastic energy (\ref{elen}), still with the same elastic constants as for the linear elastic limit.
We note that the newly defined strain-like tensor $\bar{\epsilon}_{ij}$ is rotational invariant.
For a rigid rotation (around the origin) the displacement is --- in agreement with the discussion in the preceding section --- $u_i = M_{ij} x_j$ with $M_{ij} = O_{ij} - \delta_{ij}$ with an orthogonal matrix $\mathbf{O}$, i.e.~$\mathbf{O}\mathbf{O}^\dagger=\mathbf{O}^\dagger\mathbf{O}=\mathbf{1}$ and the dagger as transposition symbol.
In coordinate notation this means $O_{ik}O_{jk}=\delta_{ij}$.
Inserting this into the expression for the nonlinear strain gives indeed $\bar{\epsilon}_{ij}=0$.
This is the expected symmetry, as the box operator was introduced to recover the rotational invariance of the amplitude equations.
Notice that in comparison the linearised strain tensor (\ref{linstrain}) is not rotational invariant.

However, from theory of elasticity we would have expected that instead of the tensor $\bar{\epsilon}_{ij}$ rather the nonlinear elastic strain tensor $\epsilon_{ij}$ should appear.
It is defined as 
\begin{equation} \label{righttensor}
\epsilon_{ij} = \frac{1}{2} \left( \frac{\partial u_i}{\partial x_j} +  \frac{\partial u_j}{\partial x_i} +  \frac{\partial u_k}{\partial x_i}  \frac{\partial u_k}{\partial x_j}\right).
\end{equation}
This tensor is also rotational invariant.
Both tensors, $\epsilon_{ij}$ and $\bar{\epsilon}_{ij}$ differ only by the nonlinear contributions.
The strain tensor $\bar{\epsilon}_{\alpha\beta}$ in (\ref{wrongtensor}) is related to the left Cauchy-Green deformation tensor $\mathbf{B} = \mathbf{F} \mathbf{F}^{\dagger}$ ($\mathbf{F}$ is the deformation gradient), i.e.~$\bar{\epsilon}_{\alpha\beta} = 1/2 (B_{\alpha\beta}-\delta_{\alpha\beta}$). 
On the other hand, the strain in (\ref{righttensor}) represents the ``conventional'' Green strain tensor, i.e., $\epsilon_{\alpha\beta}=1/2 (C_{\alpha\beta} - \delta_{\alpha\beta})$, where $\mathbf{C} = \mathbf{F}^{\dagger}\mathbf{F}$ is the right Cauchy-Green deformation tensor. 

Chan and Goldenfeld arrive at the same conclusion that the tensor $\bar{\epsilon}_{ij}$ instead of $\epsilon_{ij}$ appears in the elastic energy of a two-dimensional amplitude equations model, which is derived from the corresponding phase field crystal model with hexagonal symmetry \cite{Chan:2009aa}.
For fixed amplitudes they are able to represent the elastic energy as $F_{el}\sim \bar{\Delta}$ with
\begin{equation}
\bar{\Delta} = \frac{3}{2} \bar{\epsilon}_{xx}^2 + \frac{3}{2} \bar{\epsilon}_{yy}^2 +2 \bar{\epsilon}_{xy}^2 + \bar{\epsilon}_{xx}\bar{\epsilon}_{yy}.
\end{equation}
This expression is however identical to the same quantity defined through the conventional strain tensor,
\begin{equation}
{\Delta} = \frac{3}{2} {\epsilon}_{xx}^2 + \frac{3}{2} {\epsilon}_{yy}^2 +2 {\epsilon}_{xy}^2 + {\epsilon}_{xx}{\epsilon}_{yy},
\end{equation}
and therefore it is possible to express the elastic energy entirely through $\epsilon_{ij}$.
We note that this miraculous identity, which is not obvious on the level of the individual amplitudes, as discussed above, appears only when the summation over the set of reciprocal lattice vectors is carried out.
Surprisingly, a similar identity does {\em not} hold for the three-dimensional bcc model, and therefore it is not possible to write the elastic energy there in terms of the conventional strain tensor.
It remains therefore an open question, how nonlinear elastic deformations in the amplitude equations model relate to the standard theory of elasticity in an intuitive manner.

\section{Summary and conclusions}
\label{summary}

In this article we have investigated several phenomena related to grain boundaries dynamics using the amplitude equations model.
This model has been introduced, and its relation both to classical density functional theory and conventional phase field models has been worked out.
The amplitude equations automatically contain an appropriate description of linear elasticity.
Also, due to the description in terms of several amplitudes, which are related to the principal reciprocal lattice vectors and which serve as long-range order parameters, the preferred primary solidification in a bcc phase, which is observed for many elements, is reflected in the model.
The setup of straight symmetric tilt grain boundaries requires in spectral implementations of the amplitude equations model to satisfy periodicity conditions at the boundaries.
Here we have shown that these constraints are related to the selection of certain coincidence site lattices.
The coupling motion of a grain boundary modelled by the amplitude equations, which is subjected to shear, is well described by Cahn's and Taylor's theory in the absence of sliding at low temperatures.
Also, the phenomenon of grain rotation is captured by the amplitude equations model, and again in good agreement with theoretical predictions.

Despite all these important applications of the amplitude equations model and the benchmark against theoretical predictions, one should also keep in mind the limitations of this continuum model.
Here we have pointed out that due to the inability to describe large angle grain boundaries correctly, as the amplitude equations do not reflect properly the discrete rotation symmetry of the physical situation, also transitions between different coupling modes can be suppressed.
Also, for large elastic deformations, open questions remain with respect to the geometrical nonlinearities in the strain tensor in the amplitude equations and the related phase field crystal models.

We therefore conclude that the amplitude equations are a powerful method for large scale simulations of microstructural evolution with full atomic resolution on extended time\-scales.
Their use requires care in order to circumvent the limitations of the model in its present form.

\begin{acknowledgements}

R.S.~thanks Nigel Goldenfeld for valuable discussions concerning the geometric nonlinearities during elastic deformations.
This work has been supported by the DFG Collaborative Research Center SFB 761 {\em Steel ab initio}.

\end{acknowledgements}


\bibliographystyle{spmpsci}      

\bibliography{references}

\end{document}